\documentclass[conference]{IEEEtran}
\IEEEoverridecommandlockouts
\usepackage[caption=false,font=footnotesize]{subfig}
\usepackage{caption}
\usepackage{amsmath,amssymb,amsfonts,bm}
\usepackage{algorithm,algpseudocode}
\usepackage{graphicx}
\usepackage[english]{babel}
\usepackage{pgfplots}
\pgfplotsset{compat=newest}
\usepackage[square,numbers]{natbib}
\usepackage{xcolor}

\def\BibTeX{{\rm B\kern-.05em{\sc i\kern-.025em b}\kern-.08em
    T\kern-.1667em\lower.7ex\hbox{E}\kern-.125emX}}
\begin{document}

\title{Multi-Frame Quality Enhancement On Compressed Video Using Quantised Data of Deep Belief Networks\\
}

\author{\IEEEauthorblockN{Dionne Takudzwa Chasi }
\IEEEauthorblockA{\textit{Computer Science Division  } \\
\textit{Stellenbosch University }\\
Stellenbosch, South Africa \\
Email: dtchasi97@gmail.com}
\and
\IEEEauthorblockN{ Mkhuseli Ngxande}
\IEEEauthorblockA{\textit{Computer Science Division } \\
\textit{Stellenbosch University }\\
Stellenbosch, South Africa \\
Email: ngxandem@sun.ac.za}

}
\maketitle

\begin{abstract}
In the age of streaming and surveillance compressed video enhancement has become a problem in need of constant improvement. Here, we investigate a way of improving the Multi-Frame Quality Enhancement approach. This approach consists of making use of the frames that have the peak quality in the region to improve those that have a lower quality in that region. This approach consists of obtaining quantized data from the videos using a deep belief network. The quantized data is then fed into the MF-CNN architecture to improve the compressed video. We further investigate the impact of using a Bi-LSTM for detecting the peak quality frames. Our approach obtains better results than the first approach of the MFQE which uses an SVM for PQF detection. On the other hand, our MFQE approach does not outperform the latest version of the MQFE approach that uses a Bi-LSTM for PQF detection.
\end{abstract}


\section{Introduction}
The importance of video streaming has increased over the years as streaming has become popular over the past few years thus leading to the vast research in the field of video enhancement \cite{b22}. Video quality deteriorates during transmission due to the bandwidth, delay and loss requirements as a result videos are compressed before transmission\cite{b7}. Video compression results in deterioration of the video quality as compression artifacts are introduced in the video. Over the years, different ways have been developed for the process of video enhancement of compressed video to curb the deterioration of video quality due to video compression \cite{b4}. Video enhancement techniques can be divided into two broad categories which are self-enhancement and frame-based fusion enhancement\cite{b4}. The self-enhancement techniques do not involve embedding high quality background information into the video these techniques include  HDR-based video enhancement, compressed-based video enhancement and wavelet-based transform video enhancement\cite{b4}. On the other hand, there frame-based fusion enhancement involves extracting high quality background information and embedding this information into the low quality video to improve the quality \cite{b4}.  The approach that we will be discussing in this paper is a frame-based fusion enhancement that is done by making use of deep learning techniques. Deep Learning has been used in recent works to enhance the video quality for instance in the Decoder-side Scalable Convolutional Network (DS-CNN) where the DS-CNN learns a model to reduce the I and B/P frames in HEVC\cite{b14}. Similar to the above mentioned there is a Denoising Convolutional Convolutional Network(DnCNN) where the CNN is trained to tackle general denoising such as gaussian denoising by removing the latent clean image in the hidden layer \cite{b13}. The above mentioned methods do not make use of the frames that have good qaulity but rather they process a single frame at a time. The approach that we will discuss in this paper is the Multi Frame Quality Enhancement using Quantized data technique. Using Quantized Data from a Deep Belief Network as an input to a Bi-LSTM was used in Sleep Stage Classification\cite{b19} and in this research this approach will be applied to the Bi-LSTM that is used to detect PQF Frames.  This technique takes advantage of the fluctuating quality between the frames in the video.  In this technique the video is analysed on a frame basis. Thereafter, these frames are categorised into peak quality frames and low quality frames. The Peak Quality Frames (PQF) are used to improve the quality of the low quality frames in the neighbourhood of the PQF. Determining the PQF frames is dependent on the following video quality metrics Peak Signal to Noise Ratio (PSNR), Structural Similarity (SSIM) and the Quantization Parameter. Fig 1, fig 2 and fig 3 show how theses values change across the different frames in the video. 
\begin{figure}
\centering
\includegraphics [width=0.5\textwidth]{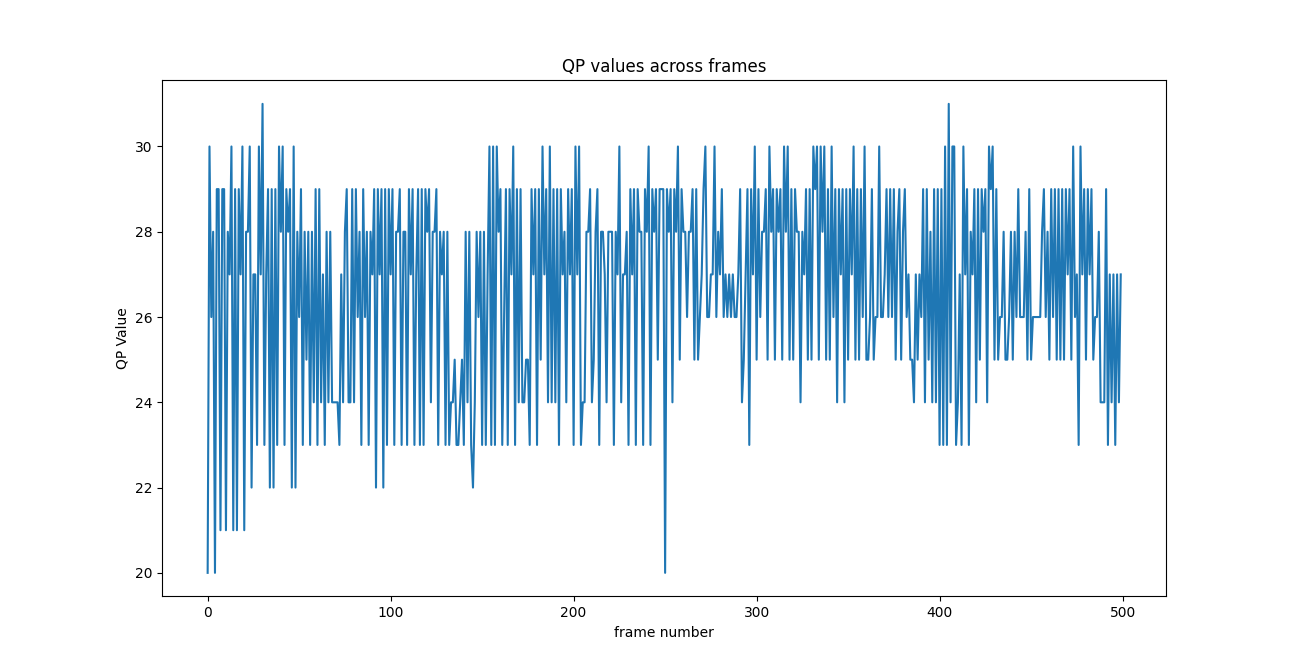}
\caption{QP Values across frames}
\end{figure} 
The QP value of a frame is related to the quality of that frame. A lower QP value relates to higher quality for that frame, it is shown that the lower QP values produce better PSNR values\cite{b17}. 
\begin{figure}
  \centering
  \includegraphics[width=0.5\textwidth]{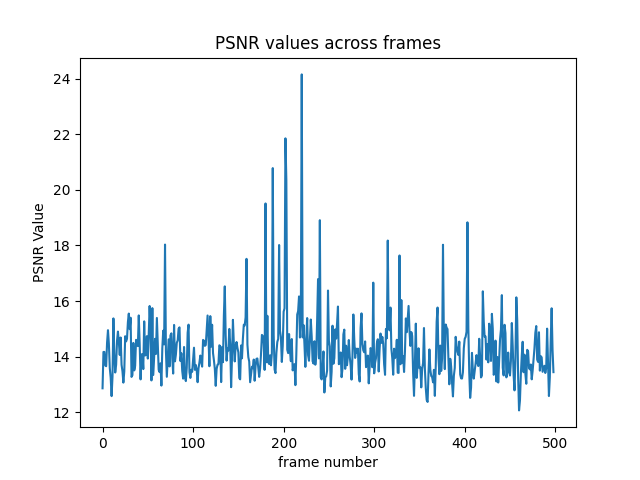}
  \caption{PSNR values across frames}
\end{figure}
The PSNR value shows the variation of video quality over the frames in the video \cite{b15}. The localized peaks are used to classify which of the frames are at peak quality. 
\begin{figure}
\centering
  \includegraphics[width=0.5\textwidth]{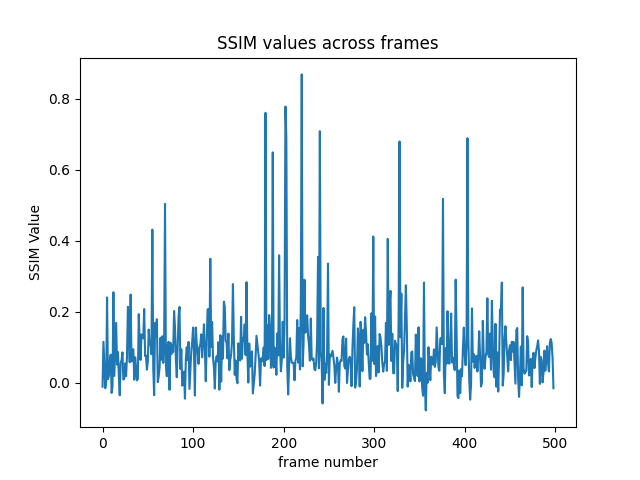}
  \caption{SSIM values across frames}
\end{figure}

The SSIM metric has been used in image restoration, video quality monitoring , image enhancement, video compression, visual, visual recognition and image coding \cite{b16}. The SSIM accounts for the visual of changes in the luminance, contrast and structure \cite{b16}. A higher structural similarity relates to better quality of a frame.

This work is an extended version of the MFQE 2.0 approach \cite{b1}. In our approach we have added another layer to the architecture by adding a DBN that quantizes the video data. This is done by using the PSNR and SSIM values for each as input to the DBN alongside labels that indicate whether the frame is a PQF or a non PQF. The quantised data will be the output of the DBN as in \cite{b19}  . The quantised data is then used for PQF detection. This approach results in faster detection of PQF frames and improves the Peak signal to noise ratio of the non-PQF's.

\section{Methodology}

\subsection{Dataset and Preprocessing }
The dataset used in this research is from Xiph.org \cite{b18} , VQEG and the Joint Collaborative Team on Video Encoding (JCT-VC). These videos contain various resolutions. These videos were then compressed to MP4 and and H.264/AVC. FFMPeg is used for the process of compressing. The quantization parameter, PSNR are associated with the quality of a frame\cite{b1}. In this implementation the metrics mentioned above are used to distinguish between PQF and non-PQF. Video Frames were extracted from the compressed video using FFMPEG, thereafter the QP value, PSNR and the SSIM were computed for each frame using OpenCV and the Skimage Library. The above mentioned values were used to create binary labels to distinguish between a PQF and a non-PQF frames.
\subsection{Quantization of Data}
 For the quantization of the data, we used a DBN . The DBN network consists of two RBM layers. The RBM layers which have 256 neurons, a learning rate of 0.05 and are trained with 100 epochs. A softmax classifier is added to the DBN. The softmax classifier outputs one-dimensional data which is then regarded as quantized data. 
 \begin{figure}
  \centering
  \includegraphics[width=0.5\textwidth]{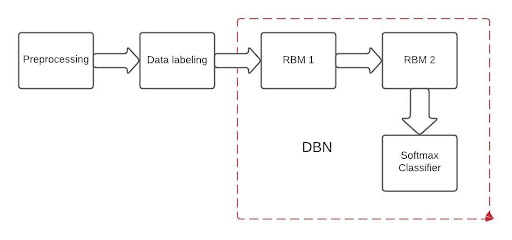}
  \caption{Quantization of Data}
\end{figure}

\subsection{PQF Detection}
The architecture of the PQF Detector is shown in Fig 5. The quantized data is taken as input by the Bi-LSTM. The Bi-LSTM is used to detect the PQF and non-PQF's based on the quality fluctuations in the video. The Bi-LSTM consists of an Embedding Layer, the bidirectional  layer that has 32 units, a dropout out layer has a dropout of 0.4(this was done to reduce overfitting within the neural network) and a dense layer with uses the ReLU activation function.

\begin{figure}
  \centering
  \includegraphics[width=0.5\textwidth]{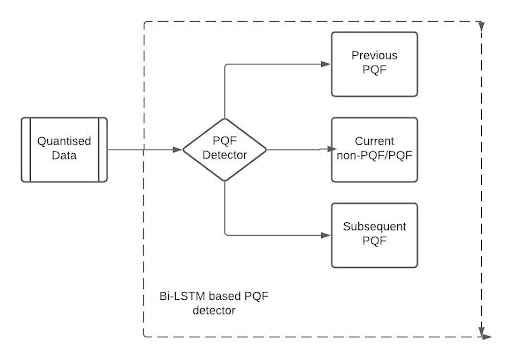}
  \caption{PQF Detection}
\end{figure}
The output of the PQF Detector is further processed such to account for the cases where there are consecutive PQF's. This is done by making use of the the probabilities of the consecutive PQF frames and selecting the PQF frame that has the highest probability as the PQF. If it is the case that they have an equal probability we then set either of them as the PQF or non PQF. These labels are further processed to account for the case where there are consecutive non-PQF's. It was stated that the average maximum separation between a non-PQF and a PQF is 2.66 \cite{b1}. Therefore if the distance between two PQF's is greater than this value one of the non-PQF's would have to act as a PQF. 

\subsection{MF-CNN}
The MF-CNN makes use of the PQF labels to improve the quality of the compressed video. The MF-CNN consists of two parts that is the MC-Subnet(Motion Compensation Subnet) and the QE-Subnet(Quality Enhancement -Subnet). 

\begin{figure}
  \centering
  \includegraphics[width=0.5\textwidth]{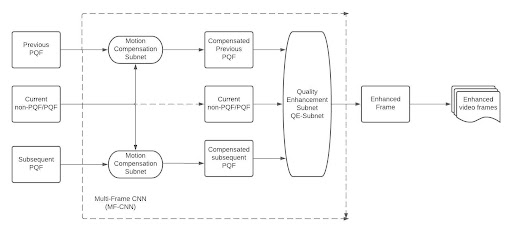}
  \caption{Multiframe Convolutional Network}
\end{figure}

\subsubsection{MC-Subnet}
The MC-Subnet is used to compensate for the differences in frames caused by temporal motion \cite{b1}. The diagram below shows the architecture of the MC-Subnet \cite{b1}.

\begin{figure}
  \centering
  \includegraphics[width=0.5\textwidth]{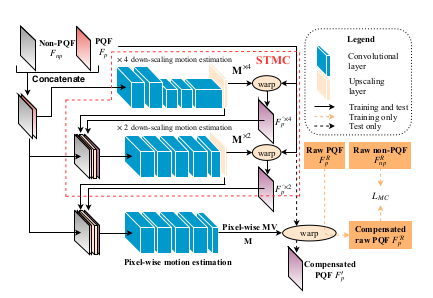}
  \caption{MC-Subnet}
\end{figure}

The MC-Subnet is responsible for compensating for the difference between PQF's and non- frames caused by the temporal motion. The MC-Subnet has  5 convolutional layers for pixel-wise motion estimation. The first 4 layers have a filter size of $ 3 \times 3 $, filter number of 24 , a stride of 1 and the PReLU activation function is used.The last layer has a filter size of $ 3 \times 3 $. The MC Subnet consists of convolutional layers that estimate the $ \times 4$ and the $ \times 2$ down scaling motion vectors. Large scale motion is handled by the down scaling motion estimation \cite{b1}. The convolutional networks that are responsible for the down scaling motion vector estimation outputs the $ \times 4 $ , $ \times 2$ maps and their corresponding PQF's The original PQF and non PQF are concatenated with this output and this is used as input to the convolutional networks for pixe wise motion estimation. The pixel wise convolutional layers output the pixel-wise motion vector which we will denote as \emph{M}. In the down scaling MV map, the horizontal MV map \emph{$M_x$} and the vertical MV map \emph{$M_y$} where x and y corresponds to the horizontal and vertical index of each pixel. The PQF is warped based on \emph{$M_x$} and \emph{$M_y$} to compensate for the temporal motion \cite{b1}. The compensated PQF will be derived using the equation :
\begin{equation}\label{eq1}
F_p^{'}(x,y) = I{F_p(x+\emph{$M_x$}(x,y),y + \emph{$M_y$}(x,y))}
\end{equation}
 where $F_p$ and $F_np$ refer to the compressed PQF and non-PQF respectively and I denotes bilinear interpolation.
 
 The MC-subnet is trained by minimizing the MSE(Mean Squared Error) between the compensated adjacent frame and the current frame \cite{b20}. Due to the use of compressed PQF's and non-PQF's there is a significant impact on the frame quality leading to reduced quality. To obtain the correct estimate of the distorted  MV, MC-subnet is trained with the supervison of the raw frames of the video \cite{b1}. This is done by warping the PQF using the MV map output and further minimizing the MSE between the compensated raw PQF and the raw non-PQF \cite{b1}.
\subsection{Quality Enhancement}
The Quality Enhancement Subnet is used to enhance the quality of the non-PQF's. The architecture of the QE-Subnet is shown below \cite{b1}.

\begin{figure}
  \centering
  \includegraphics[width=0.5\textwidth]{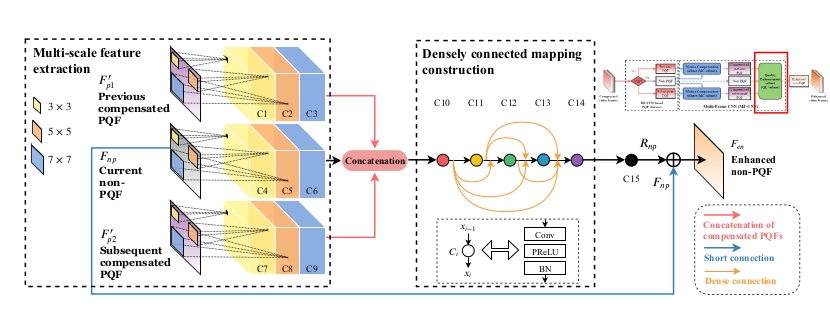}
  \caption{QE-Subnet}
\end{figure}

The non-PQF , previous compensated PQF and the subsequent compensated PQF are used as input QE-Subnet. By using the above mentioned frames the QE-Subnet takes account of the spatial and temporal features of the frames and this information is used to improve the quality of the non-PQF. The QE-Subnet consists of the multi-scale feature extraction and a densely connected mapping construction. Multi-scale convolutional filters are used extract the spatial features \cite{b1}. These convolutional layers output 288 feature maps for the input frames. The feature maps are concatenated and used as input to the densely connected mapping construction. The densely connected mapping construction is used to map the feature maps to the difference between the original frames and the enhanced frames. This is done by using five convolutional filters with a size of $3 \times 3$, batch normalization and PReLU is applied to all of the layers. The enhanced frame is computed by summing the pixels in the non-PQF $F_{en}$ and the learned difference between the original frame and the enhanced frame $D_{np} \theta_{qe}$ as shown below:
\begin{equation}\label{eq2}
F_{en} = F_{np} + D_{np} \theta_{qe}
\end{equation}

The MC-Subnet and the QE-Subnet are trained jointly \cite{b1}. The loss function of the MF-CNN is shown below:

\begin{equation}\label{eq3}
\begin{split}
L_{MF}(\theta_{mc},\theta{qe}) = a . \sum{2}^{i=1} || F_p1^{'R}(\theta_{mc}) -  F_{np}^{R}||_2^{2} + \\
b . ||( F_{np} + R_{np}(\theta_{qe})) - F_{np}^{R}||_2^{2}
\end{split}
\end{equation}

The training is divided into two steps. Initally, a>>b because $F_{p1}^{'}$ and $F_{p2}^{'}$ are generated by the MC-Subnet and are used as input into the QE-Subnet. Once the $L_{MC}$ converges,  a << b is set to minimize the MSE between $F_{np} + R_{np} and F_{np}^{R}$. Thereafter, the MF-CNN is trained to enhance the video quality.

\section{Results}

\subsection{Performance of the Deep Belief Network}
The PQF detection is highly dependent on the output of the Deep belief Network(DBN) as a result it is essential that the performance of the deep belief network is investigated. When training the deep belief network 160 videos from Xiph.org \cite{b18}. When analysing the performance of the four video sequences are used the Basketball, Mother and Daughter, Football and Coast guard video sequences are used. From table 1 it can be noted the coast guard video sequence obtains the highest accuracy and the lowest accuracy is obtained from the Mother video sequence.
\begin{table}[htbp]
\caption{Performance of Deep Belief Network }
\begin{center}
\begin{tabular}{|c|c|c|c|}
\hline
&\multicolumn{3}{|c|}{\textbf{Model Metrics }} \\
\cline{2-4} 
 \textbf{Video Sequence}& \textbf{\textit{Precision \%}}& \textbf{\textit{Recall \%}}& \textbf{\textit{$F_1$score \%}} \\
\hline
Basketball & 100 & 97.7 &  98.8\\
Mother  & 100 &  94.1 &  93.6 \\
Football & 100 &  95.4 &  97.4 \\
Coastguard & 100 &  98.0&  99.5 \\
\hline
\end{tabular}
\label{tab1}
\end{center}
\end{table}

\subsection{Performance of the Bi-LSTM PQF Detector}
The Video Quality is largely dependant on how the PQF labels are detected as a result it is vital that we investigate the performance of the PQF detector. When evaluating the performance of the PQF Detector we test make use of 160 videos. In this report we will discuss the results that were obtained from four videos that were from different scenes namely a basketball match, a mother and daughter, football match and a coastguard scene. From the graph below it is noted that the models accuracy increases against the number of epochs and the loss gradually decreases against the number of epochs. The sudden drop in the loss may be attributed to the learning rate being reaching a low at epoch 25.

\begin{figure}
  \centering
  \includegraphics[width=0.5\textwidth]{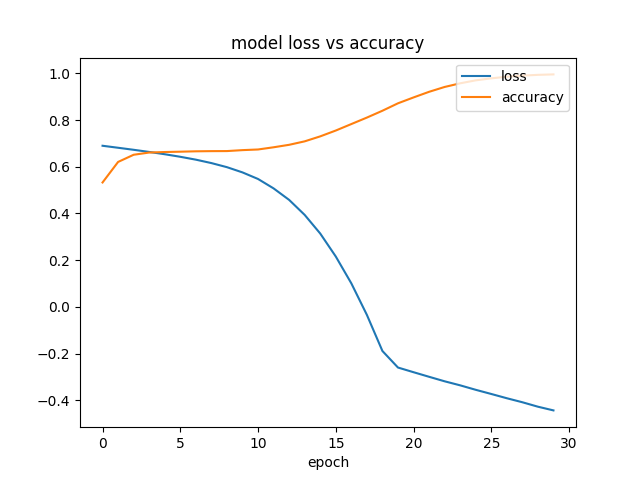}
  \caption{Model Loss vs Model Accuracy for Bi-LSTM PQF Detector }
\end{figure}

For the purpose of this report we will refer to the initial implementation of the MFQE approach as the MFQEv1\cite{b2} , the second implementation will be referred to as the MFQEv2 \cite{b1} approach and this implementation as MFQEv3. From the table below we note that the MFQEv3 is outperformed by the  MFQEv1 and the MFQEv2 approach. 

\tabcolsep=0.05cm
\begin{table}[htbp]
\caption{Performance of PQF Detector }
\begin{center}
\begin{tabular}{|c|c|c|c|c|}
\hline
&\multicolumn{4}{|c|}{\textbf{Model Metrics }} \\
\cline{2-5} 
 \textbf{Approach }&\textbf{ Video Sequence}& \textbf{\textit{Precision \%}}& \textbf{\textit{Recall \%}}& \textbf{\textit{$F_1$score \%}} \\
\hline
 MFQEv1 &Basketball & 100 & 92.4&  91.5\\
 MFQEv1 &  Mother & 100 &  96.3 & 97.2 \\
 MFQEv1 &  Football  & 100 &  89.0 & 90.2 \\
 MFQEv1 &  Coastguard & 100 &  97.4 & 98.0 \\
 MFQEv2 &Basketball & 100 & 90.0 & 93.4\\
 MFQEv2 &  Mother  & 100 &  98.4 & 99.3 \\
 MFQEv2 &  Football  & 100 & 97.0 & 98.4 \\
 MFQEv2 &  Coastguard & 100 & 97.2 & 96.5 \\
 MFQEv3 &Basketball  & 100 & 81.6 &  84.2\\
 MFQEv3 &  Mother  & 100 & 92.2 & 95.7 \\
 MFQEv3 &  Football & 100 &  87.2 & 90.1\\
 MFQEv3 &  Coastguard & 100 & 85.4 & 87.7 \\
\hline
\end{tabular}
\label{tab1}
\end{center}
\end{table}

Furthermore, it is noted that the basketball scene obtains the least accuracy and f1-score compared to the other scenes that were used in this experiment.

\subsection{Assessment of Video Quality Improvement}

From this we table we can note that the MFQEv3 is outperformed by the MFQEv2. The MFQEv3 approach outperforms the MFQEv1. It can be noted that all the approaches perform the best for the coastguard video. The MFQEv3 perfomed the best on the coastguard video (0.802dB) which is approximately a 12 \% increase when compared to the MFQEv1 approach. The MFQEv2 approach performed better than the MFQEv3 approach by approximately 7\% when related to the PSNR. The SSIM metric indicates the same pattern as the PSNR metric, the MFQEv2 outperforms both the MFQEv1 and MFQEv2 approach. It can be noted that the MFQEv3 has an SSIM that is 38\% higher than the 

\begin{table}[htbp]
\caption{Overall comparison of PSNR and SSIM}
\begin{center}
\begin{tabular}{|c|c|c|c|}
\hline
&\multicolumn{3}{|c|}{\textbf{Video Metrics }} \\
\cline{2-4} 
 \textbf{Approach }&\textbf{ Video Sequence}& \textbf{\textit{ $\delta$ PSNR}}& \textbf{\textit{SSIM}} \\
\hline
 MFQEv1 &Basketball Scene & 0.658 & 81.6 \\
 MFQEv1 &  Mother and Daughter & 0.610 & 92.2 \\
 MFQEv1 &  Football Match & 0.459 &  87.2 \\
 MFQEv1 &  Coastguard & 0.798 & 100  \\
 MFQEv2 &Basketball Scene & 0.728 & 115\\
 MFQEv2 &  Mother and Daughter & 0.585 &  102 \\
 MFQEv2 &  Football Match & 0.516 & 83  \\
 MFQEv2 &  Coastguard & 0.960 & 157  \\
 MFQEv3 & Basketball Scene & 0.628 & 97.4\\
 MFQEv3 &  Mother and Daughter & 0.477 &  90  \\
 MFQEv3 &  Football Match & 0.406 &  80  \\
 MFQEv3 & Coastguard & 0.892 &  138 \\
\hline
\end{tabular}
\label{tab1}
\end{center}
\end{table}

\begin{figure}
  \centering
  \includegraphics[width=0.5\textwidth]{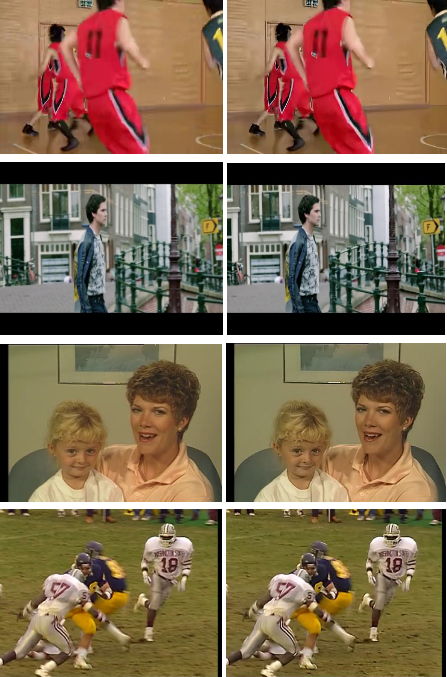}
  \caption{Model Loss vs Model Accuracy for Bi-LSTM PQF Detector }
\end{figure}

Figure 10 show images above where the compression artefacts are reduced slightly in the second image. It can be noted that the coastguard video has the best quality before and after enhancement. By taking a closer look at the football match video it can be seen that there is a reduced amount of artefacts especially around the legs of the number 57 in the video. In the mother and daughter frame it can be seen that there is a reduction in artefacts around the daughters hair.

\section{Discussion}
As shown above it can be seen that the performance of the MFQEv3 has been compared to the MFQEv1 and the MFQEv2. In this paper we have assessed the performance of these implementations based on the frame enhancement and performance of the PQF detector. The results show that all approaches perform the best on the coast guard video. This may attributed to the coast guard video generally having a higher resolution as compared to the other videos. It is import to note that there are other aspects that impact video improvement that is not accounted for in this report. Aspects such as bit-rate, bit-depth and bit control mode may affect to what extent the quality of a video may be improved. A higher bit-rate may be attributed to better video quality \cite{b21}. Although the bit-rate has an impact on quality there is a point the video quality reaches a point where increasing the bit-rate will only result in a bigger file size without file enhancement. The diagram below describes the relationship between bit-rate, frame rate, resolution and the video quality.

\begin{figure}
  \centering
  \includegraphics[width=0.5\textwidth]{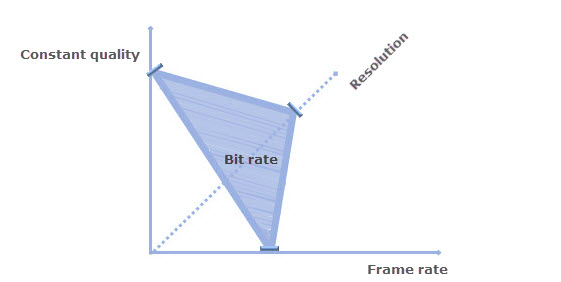}
  \caption{Relationship between bit-rate, frame-rate, resolution and video quality}
\end{figure}

All these factors listed above can contribute to which videos obtain the highest quality enhancement. These factors can be used to explain the football match video being the one to obtain the least amount of video enhancement. The bit depth of the video is an attribute that affects the quality improvement of a video. The bit depth refers to the number of  bits used for each color component of a single pixel. Thus, if a frame has a high bit depth, the frame will be finer and smoother \cite{b21}. As seen in the images above the basketball video is seen to have a few artifact thus indicating a low bit-depth. A frame that has a low bit-rate is more difficult to improve the quality because of the increased artefacts.

It is noted that the accuracy of the PQF detector rises constantly and converges at an accuracy of 0.90. On the other hand the model loss of the PQF detector is decreases constantly and there is a sudden drop in the loss at epoch 17 thereafter the loss decreases linearly. This may be  due to the learning reaching a point where it saturates the learning rate. This is when each step update keeps jumping across the minima.

\section{Conclusion}
In this paper, the subject of this research is the performance of using quantized data in the MFQE approach for video enhancement to reduce the compression artefacts. This differs from the MFQEv1 and MFQEv2 where the video data directly to improve detect the PQF's. In our approach we first parse this data through a deep belief network that has two RBM layer and a softmax classifier at the end. This was regarded as our quantized data following the approach from \cite{b19}. This quantized data was fed into the Bi-LSTM PQF detector where we the video frames where classified as non-PQF or PQF's. The PQF's were then parsed into the MF-CNN were the PQF's were used to enhance the non-PQF frames. The MC-subnet would then account for the motion between the non-PQF's and PQF. The QE-Subnet would then make use of the PQF's to improve the non-PQF frames. Experiments were done to compare our approach with the two implementations of the MFQE approach respectively the MFQEv1 \cite{b2} and the MFQEv2 \cite{b1}. The experiments have shown that all three approaches have resulted in overall quality enhancement. It was noted that our version of the MFQE approach outperforms the MFQEv1 but it is outperformed by the MFQEv2 approach. For the purpose of future work, it may be necessary to investigte the performance of the MFQE approach for uncompressed video and make use of a vast number of video quality metrics to decide which ones frames are PQF and non-PQF.
\section*{Acknowledgment}

I would like to thank Dr Xing, Department of Electronics Information Engineering, Beihang University for his comments and advice during the course of this research. I would also like to thank the videographers involved in uploading videos on the Xiph.org. I would also like to thank my supervisor Dr Ngxande, Department of Computer Science, Stellenbosch University for his continual assistance with this work. I would like to thank my family and friends for their support throughout the course of this research.


\begin{thebibliography}{00}
\bibitem{b1} Guan, Z., Xing, Q., Xu, M., Yang, R., Liu, T. and Wang, Z., 2021. MFQE 2.0: A New Approach for Multi-Frame Quality Enhancement on Compressed Video. IEEE Transactions on Pattern Analysis and Machine Intelligence, 43(3), pp.949-963.
\bibitem{b2} Xing, Q., Guan, Z. and Sneddon, I., 2021. Multi-frame Quality Enhancement on Compressed Video. CVF Conference on Computer Vision and Pattern Recognition (CVPR).
\bibitem{b3} Ou, Y., Ma, Z., Liu, T. and Wang, Y., 2011. Perceptual Quality Assessment of Video Considering Both Frame Rate and Quantization Artifacts. IEEE Transactions on Circuits and Systems for Video Technology, 21(3), pp.286-298.
\bibitem{b4}Elissa, K., n.d. A survey of Video Enhancement Techniques.
\bibitem{b6} Yorozu, Y., Hirano, M., Oka, K. and Tagawa, Y., 1987. Real-Time transmission of video streaming video streaming over computer Networks. IEEE Translation Journal on Magnetics in Japan, 2, pp.740-741.
\bibitem{b7} Yorozu, Y., Hirano, M., Oka, K. and Tagawa, T., 1987. Streaming video over the Internet: Approaches and directions. IEEE Translation Journal on Magnetics in Japan, 2, pp.740-741.
\bibitem{b8}Liew, A. and Yan, H., 2004. Blocking Artifacts Suppression in Block-Coded Images Using Overcomplete Wavelet Representation. IEEE Transactions on Circuits and Systems for Video Technology, 14(4), pp.450-461.
\bibitem{b9} Foi, A., Katkovnik, V. and Egiazarian, K., 2007. Pointwise Shape-Adaptive DCT for High-Quality Denoising and Deblocking of Grayscale and Color Images. IEEE Transactions on Image Processing, 16(5), pp.1395-1411.
\bibitem{b10} Wang, C., Zhou, J. and Liu, S., 2013. Adaptive non-local means filter for image deblocking. Signal Processing: Image Communication, 28(5), pp.522-530.
\bibitem{b11} Wang, C., Zhou, J. and Liu, S., 2013. Adaptive non-local means filter for image deblocking. Signal Processing: Image Communication, 28(5), pp.522-530.
\bibitem{b12} Janscary, J., Nowozin, S. and Rother, C., 2012. Loss-specific training of non-parametric image restoration models: A new state of the art. Proceedings of the European Conference on Computer Vision, pp.112-125.
\bibitem{b13} Jung, C., Jiao, L., Qi, H. and Sun, T., 2012. Image deblocking via sparse representation. Signal Processing: Image Communication, 27(6), pp.663-677.
\bibitem{b13} Yang, R., Xu, M. and Wang, Z., 2007. Decoder-side HEVC quality enhancement with scalable convolutional neural network. International Congress on Mathematical Education:, pp.817-822.
\bibitem{b14} Zhang, K., Zuo, W., Chen, Y., Meng, D. and Zhang, L., 2017. Beyond a Gaussian Denoiser: Residual Learning of Deep CNN for Image Denoising. IEEE Transactions on Image Processing, 26(7), pp.3142-3155.
\bibitem{b15}Huynh-Thu, Q. and Ghanbari, M., 2010. The accuracy of PSNR in predicting video quality for different video scenes and frame rates. Telecommunication Systems, 49(1), pp.35-48. 
\bibitem{b16}Dosselmann, R. and Yang, X., 2009. A comprehensive assessment of the structural similarity index. Signal, Image and Video Processing, 5(1), pp.81-91.
\bibitem{b16}Wu, C., Su, P., Huang, L. and Chiou, C., 2013. Constant frame quality control for H.264/AVC. APSIPA Transactions on Signal and Information Processing, 2.
\bibitem{b17}Balaji, L., n.d. Performance analysis of quantization parameter in base and enhancement layer in scalable video coding.
\bibitem{b18}Media.xiph.org. 2021. Xiph.org \:\: Derf's Test Media Collection. [online] Available at: <https://media.xiph.org/video/derf/> [Accessed 4 April 2021].
\bibitem{b19}Yulita, I., Fanany, M. and Arymuthy, A., 2017. Bi-directional Long Short-Term Memory using Quantized data of Deep Belief Networks for Sleep Stage Classification. Procedia Computer Science, 116, pp.530-538.
\bibitem{b20}Caballero, J., Ledig, C., Aitken, A., Acosta, A., Totz, J., Wang, Z. and Shi, W., 2017. Real-time video super-resolution with spatio-temporal networks and motion compensation. Proceedings of the IEEE Conference on Computer Vision and Pattern Recognition.
\bibitem{b21}Joskowicz, J. and Carlos Lopez Ardao, J., n.d. Combining the effects of frame rate, bit rate, display size and video content in a parametric video quality model.
\bibitem{b22} 2021. [online]Available at https://cyrekdigital.com/uploads/content/files/white-paper-c11-741490.pdf [Accessed 1 October 2021].
\bibitem{b23}Hinton, G., Osindero, S. and Teh, Y., 2006. A Fast Learning Algorithm for Deep Belief Nets. Neural Computation, 18(7), pp.1527-1554.
\end{thebibliography}

\end{document}